\documentclass[prd, preprint, nofootinbib]{revtex4}

\usepackage{graphicx}
\usepackage{amsfonts}
\usepackage{latexsym}
\usepackage{amsmath}
\usepackage{amssymb}
\usepackage{epsfig}

\begin{document}


\title{Intergenerational gauged $B-L$ model and its implication to muon $g-2$ anomaly and thermal dark matter}

\author{Nobuchika Okada}
 \email{okadan@ua.edu}
 \affiliation{
Department of Physics and Astronomy, 
University of Alabama, Tuscaloosa, Alabama 35487, USA
}

\author{Osamu Seto}
 \email{seto@particle.sci.hokudai.ac.jp}
 \affiliation{Department of Physics, Hokkaido University, Sapporo 060-0810, Japan}

%

\begin{abstract}
We study the flavor dependent $U(1)_{B_i-L_j}$ models, where an $i$th generation of quarks and $j(\neq i)$th generation of leptons are charged.
By solving the anomaly free condition for the matter sector of the SM fermions and three generations of RH neutrinos,
 we find that the $j$th generation of right-handed (RH) neutrino is not necessarily charged under the $U(1)_{B_i-L_j}$ gauge symmetry with the charge $-1$ and the other (neither $i$th nor $j$th) generation of RH neutrino can also be.
As a general solution for the anomaly cancellation conditions, the other two RN neutrinos than the charge $-1$ RH neutrino may have nonvanishing charge and be stable due to the gauge invariance, and hence it is a candidate for dark matter (DM) in our Universe.
We apply this result to a $B_3-L_2$ model and consider a light thermal DM and a solution to the muon $g-2$ anomaly.
We identify the parameter region to have the DM mass range from MeV to sub-GeV and simultaneously solve the muon $g-2$ anomaly.
We also derive the constraints on the gauge kinetic mixing parameter by using the latest Borexino phase-II data.
\end{abstract}

\preprint{EPHOU-23-015} 


\maketitle


\section{Introduction}

Introduction of an extra $U(1)$ gauge interaction is one of the promising and well-defined extensions of the standard model (SM) of particle physics.
The $B-L$ (baryon number minus lepton number) appears to be an accidental
 global symmetry in the SM, indicating that this might be a gauge symmetry in a ultraviolet (UV) completion of the theory~\cite{Pati:1973uk,Davidson:1978pm,Mohapatra:1980qe,Mohapatra:1980}.
For such the extended gauge group $G=SU(3)_C\times SU(2)_L\times U(1)_Y \times U(1)_{B-L}$, 
the cancellation condition for gauge and mixed gauge-gravitational anomaly 
requires that the number of right-handed (RH) neutrinos are three as other SM fermions. 
We note that the anomaly cancellation of $U(1)_{B-L}$ gauge symmetry
 can be realized for each generation of fermions.
Thus, even if the gauge charge are generation (flavor) dependent, theories are anomaly free.
A simple example is the so-called $(B-L)_3$ model~\cite{Babu:2017olk,Alonso:2017uky,Bian:2017rpg,Cox:2017rgn,Elahi:2019drj,Okada:2019sbb} in which only the third generation are charged, and experimental bounds with the third generation fermions are relatively weak compared to those with the first and the second generation fermions.

Anyway, once we abandon the flavor universality of gauge interactions, we can consider various extra gauged $U(1)_X$ models with $X$ being a corresponding quantum number that fulfills the anomaly free conditions.
It is even possible for the anomaly to be canceled with only leptons. The total anomalies are canceled between generations in such a leptophilic gauge interaction based on the $U(1)_{L_i-Lj}$ gauge symmetry with $i$ and $j(\neq i)$ are generation indices, and among them the $U(1)_{L_\mu-L_\tau}$ gauge symmetry~\cite{He:1990pn,Foot:1990mn} has received particular attention because it can reconcile the discrepancy of the muon anomalous magnetic moment $g_{\mu}$ (muon $g-2$) between the SM prediction and experimental results~\cite{Baek:2001kca,Ma:2001md}. $B-3L_i$~\cite{Chun:2018ibr,Greljo:2021xmg,Barman:2019aku,Wang:2019byi,Barman:2021yaz} corresponds to quark flavor universality, but lepton flavor dependent charge assignment.
Other examples include $B_3-L_1$ or $B_3+(L_1-L_2-L_3)$~\cite{Crivellin:2015lwa,Bonilla:2017lsq,Ko:2017quv}, where the total anomaly is also canceled between generations. 

In this paper, we study the flavor dependent $U(1)_{B_i-L_j}$ models where
 an $i$th generation of quarks and $j(\neq i)$th generation of leptons are charged under $U(1)_{B-L}$~\cite{Chun:2018ibr}.
A $B_3-L_2$ model had been studied~\cite{Allanach:2020kss,Allanach:2022blr,Ban:2021tos} in the context of the so-called $\mathcal{R}(K), \mathcal{R}(K^*)$ anomaly, but the most recent analysis has shown that the experimental results are consistent with the SM predictions~\cite{LHCb:2022qnv}.
We examine the anomaly free condition for the matter sector of the SM fermions and three generations of RH neutrinos.
We point out that the $j$th generation of RH neutrino is not necessarily charged under the $U(1)_{B_i-L_j}$ gauge symmetry with the charge $-1$ as been assigned in Refs.~\cite{Chun:2018ibr,Allanach:2020kss} and the other (neither $i$th nor $j$th) generation of RH neutrino can also be.
As a general solution for the anomaly cancellation conditions,
 the other two RH neutrinos than the charge $-1$ may have nonvanishing charge.
In such cases, the usual neutrino Dirac mass term between left-handed (LH) and RH neutrinos
 cannot be formed due to the gauge invariance,
 while the two RH neutrinos can form a Dirac fermion. This Dirac fermion is stable due to the gauge invariance unless another $U(1)_{B_i-L_j}$  charged Higgs doublets are introduced, thus it is a candidate for dark matter (DM) in our Universe.
After general discussion, we focus on $B_3-L_2$ model because it could solve the the muon $g-2$ problem~\cite{Aoyama:2020ynm}. 
We identify the parameter region to solve the muon $g-2$ problem and to realize viable light thermal weakly interacting massive particle (WIMP) with the mass in the range of MeV to sub-GeV.
Since the mediator $Z'$ boson does not couple light quarks and the DM mass is small, this Dirac DM is free from the constraints of direct DM search experiments. 
 
This paper is organized as follows. In the next section, 
 we examine the anomaly free condition without introducing extra fermions except three RH neutrinos
  in $U(1)_{B_i-L_j}$ gauge symmetry.
After Sec.~\ref{Sec:g-2}, we focus on $U(1)_{B_3-L_2}$ model. 
We provide the formula for the muon $g-2$ in Sec.~\ref{Sec:g-2},
 and show the favored parameter region of light thermal WIMP and to solve the muon $g-2$ problem at the same time in Sec.~\ref{Sec:DM}.
In Sec.~\ref{Sec:Constraints}, we derive the constraints on the gauge kinetic mixing by examining the electron neutrino scattering experiments.  
Sec.~\ref{Sec:Summary} is devoted to summary.

\section{$B_i-L_j$ Model}
\label{Sec:BiLj}

As a variant of $U(1)_{(B-L)_i}$ gauge symmetry, we consider cases where different generations of quarks and leptons are charged under the extra $U(1)_X$ gauge symmetry.
The total gauge symmetry is based on
 the gauge group $SU(3)_C \times SU(2)_L \times U(1)_Y \times U(1)_{B_{i}-L_{j}}$~\cite{Chun:2018ibr}. 
Anomaly free conditions can be fulfilled without introducing new fermions
 besides three RH neutrinos.
By solving the set of anomaly cancellation conditions, we find two different solutions with two free real parameters $x_H$ and $x_N$.  
The $U(1)_{B_3-L_2}$ charge assignment is shown in Tables~\ref{tableBL322} and \ref{tableBL321}.
Here, we have two cases: one is that the second generation of RH neutrino has the charge $-1$ as in Table~\ref{tableBL322}, the other is that the first of RH neutrino has the charge $-1$ as in Tabble~\ref{tableBL321}. 
Somewhat nontrivial fact is the absence of the case that $i$th generation of RH neutrino $\nu_{R}^i$ has the charge $-1$.
In both cases, the remaining other two RH neutrinos may have nonvanishing opposite charge $x_N$ and $-x_N$ each other\footnote{The similar solution had been found in a $B_3-3L_2$ model too, however only the $x_N=0$ case was investigated~\cite{Bonilla:2017lsq}.}. 
This opposite charge offers the possibility that this pair may compose a Dirac fermion, that we consider in the rest of this paper. 
Another parameter $x_H$ denotes the mixing between $U(1)_{B_i-L_j}$ and $U(1)_Y$ as the extra $U(1)$ gauge symmetry
 can be a linear combination of those two gauge symmetries~\cite{Appelquist:2002mw,Oda:2015gna,Das:2016zue}. 
Since our main interest is $x_N$ in this paper, from now on we take $x_H=0$, for simplicity.

%
\begin{table}[htb]
\begin{center}
\begin{tabular}{|c|ccc|}
\hline
 Field and representation under &  & Generation  &   \\ 
  $(SU(3)_C,SU(2)_L,U(1)_Y) $      & 1 & 2 & 3  \\ 
\hline
$Q^{k} {}_{(\mathbf{3},\mathbf{2},1/6)}$     & $0 $ &  $0 $ &  $\frac{1}{6}x_H+\frac{1}{3} $ \\
$u^{k}_{R} {}_{(\mathbf{3},\mathbf{1}, 2/3)}$& $0 $ &  $0 $ &  $\frac{2}{3}x_H+\frac{1}{3}$ \\
$d^{k}_{R} {}_{(\mathbf{3},\mathbf{1},-1/3)}$& $0 $ &   $0 $ &  $ -\frac{1}{3}x_H+\frac{1}{3}$  \\
\hline
$L^{k} {}_{(\mathbf{ 1 },\mathbf{2}, -1/2)}$ &  $0 $ &   $-\frac{1}{2}x_H-1 $&  $0 $   \\
$e^{k}_{R} {}_{(\mathbf{1},\mathbf{1}, -1)}$ &  $0 $ &  $-x_H-1$ &  $0 $  \\
$\nu^{k}_{R}{}_{(\mathbf{1},\mathbf{1}, 0)}$ & $-x_N$ & $-1$ &  $ x_N$  \\   
\hline
\end{tabular}
\end{center}
\caption{
An anomaly free $U(1)_{B_3-L_2}$ charge assignment for SM particles and RH neutrinos. 
}
\label{tableBL322}
\end{table}
%
\begin{table}[htb]
\begin{center}
\begin{tabular}{|c|ccc|}
\hline
 Field and representation under &  & Generation  &   \\ 
  $(SU(3)_C,SU(2)_L,U(1)_Y) $      & 1 & 2 & 3  \\ 
\hline
$Q^{k} {}_{(\mathbf{3},\mathbf{2},1/6)}$     & $0 $ &  $0 $ &  $\frac{1}{6}x_H+\frac{1}{3}$ \\
$u^{k}_{R} {}_{(\mathbf{3},\mathbf{1}, 2/3)}$& $0 $ &  $0 $ &  $\frac{2}{3}x_H+\frac{1}{3} $ \\
$d^{k}_{R} {}_{(\mathbf{3},\mathbf{1},-1/3)}$& $0 $ &   $0 $ &  $-\frac{1}{3}x_H+\frac{1}{3} $  \\
\hline
$L^{k} {}_{(\mathbf{ 1 },\mathbf{2}, -1/2)}$ &  $0 $ &   $-\frac{1}{2}x_H-1 $&  $0 $   \\
$e^{k}_{R} {}_{(\mathbf{1},\mathbf{1}, -1)}$ &  $0 $ &  $-x_H-1 $ &  $0 $  \\
$\nu^{i}_{R}{}_{(\mathbf{1},\mathbf{1}, 0)}$ & $-1$ & $-x_N$ & $ x_N$  \\   
\hline
\end{tabular}
\end{center}
\caption{
An anomaly free $U(1)_{B_3-L_2}$ charge assignment for SM particles and RH neutrinos. 
}
\label{tableBL321}
\end{table}

So far, we have solved the anomaly cancellation conditions of the gauge group
 $SU(3)_C \times SU(2)_L \times U(1)_Y \times U(1)_{B_{i}-L_{j}}$
  with the particle content of the SM fermions plus three RH neutrinos.
On the other hand, the solution we have obtained is in fact same as a model constructed by
 the gauged $B-L$ extension to only one generation as $U(1)_{B_i-L_j}$, where one RH neutrino is introduced for the anomaly cancellation while the other two RH neutrinos are additionally introduced not necessary for the anomaly cancellation.
Here, the $x_N=0$ case corresponds to introducing Majorana fermions, the $x_N \neq 0$ case does to introducing one vectorlike fermion.

\subsection{Gauge sector}

The gauge kinetic terms of our model are
\begin{align}
\mathcal{L}_{\mathrm{gauge}} =& -\frac{1}{4}G_{\mu\nu}G^{\mu\nu} -\frac{1}{4}\hat{W}_{\mu\nu}\hat{W}^{\mu\nu} -\frac{1}{4}\hat{B}_{\mu\nu}\hat{B}^{\mu\nu} -\frac{1}{4} \hat{X}_{\mu\nu}\hat{X}^{\mu\nu} +\frac{\sin\epsilon}{2}\hat{B}_{\mu\nu}\hat{X}^{\mu\nu},
\end{align}
where fields with the hat stand for those in gauge eigenstate and $\epsilon$ is the gauge kinetic mixing parameter. Here and hereafter, we use the symbol $X$ as the extra gauge charge and field, for simplicity.
We assume that the SM Higgs field $\Phi$ is singlet under the $U(1)_X$ and the mass of $\hat{X}$ gauge boson $M_{\hat{X}}$ is
 generated by another scalar field. At the electroweak (EW) breaking vacuum, we have
\begin{align}
\hat{A}_{\mu} &= s_W \hat{W}^3_{\mu} + c_W \hat{B}_{\mu} , \\
\hat{Z}_{\mu} &= c_W \hat{W}^3_{\mu} - s_W \hat{B}_{\mu} ,
\end{align}
with $s_W = \sin\theta_W$ and $c_W = \cos\theta_W$, where $\theta_W$ is the Weinberg angle.
The field redefinition by an orthogonal matrix, 
\begin{equation}
U_K = \left(
\begin{array}{ccc}
1 & 0 & t_{\epsilon} c_W \\
0 & 1 & -t_{\epsilon} s_W \\
0 & 0 & \frac{1}{c_{\epsilon}} \\
\end{array}
\right) ,
\end{equation}
 resolves the kinetic mixing but induces the mass mixing :
\begin{equation}
M^2 = \left(
\begin{array}{ccc}
0 & 0 & 0 \\
0 & M_{\hat{Z}}^2 & -t_{\epsilon} s_W M_{\hat{Z}}^2 \\
0 & -t_{\epsilon} s_W M_{\hat{Z}}^2 & \frac{1}{c_{\epsilon}^2}M_{\hat{X}}^2 +\left(t_{\epsilon} s_W\right)^2  M_{\hat{Z}}^2\\
\end{array}
\right) ,
\end{equation}
 with $M_{\hat{Z}}^2 = \frac{1}{4}(g_1^2+g_2^2)v^2$, $c_\epsilon = \cos\epsilon$, $t_\epsilon = \tan\epsilon$, and $M_{\hat{Z}}^2$ being the mass generated by the vacuum expectation value (VEV) of the extra $U(1)_{B_i-L_j}$ breaking scalar field.
The additional field redefinition to the mass eigenstates can be done with a rotation matrix
\begin{equation}
U_M = \left(
\begin{array}{ccc}
1 & 0 & 0 \\
0 & \cos\theta & -\sin\theta \\
0 & \sin\theta & \cos\theta \\
\end{array}
\right) ,
\end{equation}
 with the angle\footnote{The sign of $\theta$ is opposite to that in Ref.~\cite{Cho:2020mnc}.}
\begin{equation}
\tan 2\theta = \frac{-2t_{\epsilon}s_W M_{\hat{Z}}^2 }
{ M_{\hat{Z}}^2- \left(t_{\epsilon} s_W\right)^2 M_{\hat{Z}}^2 -\frac{1}{c_{\epsilon}^2}M_{\hat{X}}^2 } .
\end{equation}
The mass eigenvalues are given by
\begin{align}
m_Z^2 &= M_{\hat{Z}}^2 \cos{}^2\theta - t_{\epsilon} s_W M_{\hat{Z}}^2\sin(2\theta)+\left( \frac{1}{c_{\epsilon}^2}M_{\hat{X}}^2  + \left(t_{\epsilon} s_W\right)^2  M_{\hat{Z}}^2 \right) \sin{}^2\theta, \\
m_{Z'}^2 &= \left( \frac{1}{c_{\epsilon}^2}M_{\hat{X}}^2  + \left(t_{\epsilon} s_W\right)^2 M_{\hat{Z}}^2 \right) \cos{}^2\theta + M_{\hat{Z}}^2 \sin{}^2\theta + t_{\epsilon} s_W M_{\hat{Z}}^2\sin(2\theta) .
\end{align}
Since the hatted field and the unhatted field are related as
\begin{align} 
\left(
\begin{array}{c}
\hat{A}_{\mu} \\
\hat{Z}_{\mu} \\
\hat{X}_{\mu} \\
\end{array}
\right) = U_K U_M
\left(
\begin{array}{c}
A_{\mu} \\
Z_{\mu} \\
Z'_{\mu} \\
\end{array}
\right) 
=
\left(
\begin{array}{ccc}
 1 & c_W s_{\theta} t_{\epsilon} & c_{\theta} c_W t_{\epsilon} \\
 0 & c_{\theta}-s_{\theta} s_W t_{\epsilon} & -c_{\theta} s_W t_{\epsilon}-s_{\theta} \\
 0 & \frac{s_{\theta}}{c_{\epsilon}} & \frac{c_{\theta}}{c_{\epsilon}} \\
\end{array}
\right)
\left(
\begin{array}{c}
A_{\mu} \\
Z_{\mu} \\
Z'_{\mu} \\
\end{array}
\right) 
,
\end{align}
by combining with
\begin{align}
\left(
\begin{array}{c}
\hat{W}^3_{\mu} \\
\hat{B}_{\mu} \\
\hat{X}_{\mu} \\
\end{array}
\right) &= \left(
\begin{array}{ccc}
s_W & c_W & 0  \\
c_W & -s_W  & 0  \\
0 & 0 & 1 \\
\end{array}
\right)  
\left(
\begin{array}{c}
\hat{A}_{\mu} \\
\hat{Z}_{\mu} \\
\hat{X}_{\mu} \\
\end{array}
\right) ,
\end{align}
 we find
\begin{align}
\left(
\begin{array}{c}
\hat{W}^3_{\mu} \\
\hat{B}_{\mu} \\
\hat{X}_{\mu} \\
\end{array}
\right) 
 = \left(
\begin{array}{ccc}
s_W & c_W c_{\theta} & - c_W s_{\theta}  \\
c_W & -s_W c_{\theta} + t_{\epsilon}s_{\theta}  & s_W s_{\theta} + t_{\epsilon} c_{\theta}  \\
0 & \frac{1}{c_{\epsilon}}s_{\theta} & \frac{1}{c_{\epsilon}} c_{\theta} \\
\end{array}
\right)  
\left(
\begin{array}{c}
A_{\mu} \\
Z_{\mu} \\
Z'_{\mu} \\
\end{array}
\right) .
\end{align}

\subsection{Fermion masses and Higgs sector}

In a ``flavored'' $B-L$ symmetry as in $U(1)_{(B-L)_3}$, due to $U(1)_{B_i-L_j}$ gauge symmetry,
 the $U(1)_{B_i-L_j}$ singlet SM Higgs field $\Phi$ cannot give the masses of
 the $U(1)_{B_i-L_j}$ charged fermions
 and thus realistic fermion flavor mixings cannot be reproduced.

\subsubsection{Quark mass and mixing}

To reproduce the realistic quark mass matrices,
 a few successful UV completions for $U(1)_{(B-L)_3}$ have been proposed:
One is an extension of Higgs sector by Babu \textit{et al}. in Ref.~\cite{Babu:2017olk} 
 and another is introduction of heavy vectorlike fermions with additional scalars
 by Alonso \textit{et al}. in Ref.~\cite{Alonso:2017uky}. 
The same mechanism work for at least $U(1)_{B_3-L_j}$. 
Since the details of those UV completions are irrelevant for the following discussion,
 we will not consider a specific model further.

\subsubsection{Lepton mass and mixing}

The charged lepton masses can be generated by the SM Higgs field, if $x_H=0$.
On the other hand, if $x_H \neq 0$ , we need to introduce 
 the second Higgs doublet with $(\mathbf{ 1 },\mathbf{2}, -1/2, x_H/2)$
 to generate all charged lepton masses.
The generation of neutrino mass also depends on the Higgs sector.
Here we consider only the $x_H=0$ case for simplicity.

The $x_N=0$ case is simple. Since two RH neutrinos are singlet under any gauge group in this case,
 the type-I seesaw mechanism~\cite{Minkowski:1977sc,Yanagida:1979as,GellMann:1980vs,Mohapatra:1979ia} works through neutrino Yukawa coupling with the SM Higgs field $\Phi$ and their Majorana mass. 

For $x_N \neq 0$ , on the other hand, we need to extend the model in order to generate observed neutrino masses. The simplest extension would be introducing $SU(2)$ triplet Higgs fields
 $\Delta$~\cite{Schechter:1980gr,Magg:1980ut,Cheng:1980qt}, since the Dirac neutrino mass between 
 the extra $U(1)$ charged RH neutrinos and the extra $U(1)$ uncharged LH neutrinos are not necessary.
The Yukawa couplings are given by
\begin{align}
 \mathcal{L}_\mathrm{Yukawa} \supset 
   & + \sum_{k,l}\left(-\frac{1}{\sqrt{2}}y^{\Delta_0}_{kl}\overline{L^{k~C}}\cdot \Delta_0 L^l
     -\frac{1}{\sqrt{2}}y^{\Delta_1}_{kj}\overline{L^{k~C}}\cdot \Delta_1 L^j \right)    \nonumber  \\
  &   - \sum_{k,l} y^{e}_{kl}\overline{L^k} \Phi e_R^l - y^{D}_{j{j'}} \overline{L^j} \widetilde{\Phi} \nu_R^{j'} - \frac{1}{2}y^{\nu_R}_{j'} \overline{\nu_R^{j'~C}} \phi_2 \nu_R^{j'} + \mathrm{ H.c.} ,
\label{Lag1} 
\end{align}
 where the superscript $C$ denotes the charge conjugation, the dot denotes the antisymmetric product of $SU(2)$, 
 subscripts of $\Delta$ denote these $U(1)_X$ charges,
 and $k$ and $l$ run from $1$ to $3$ except the $j$th generation. 
$\nu_R^{j'}$ denotes the RH neutrino with the charge $-1$. 
For instance, $j'=2$ in the model in Table~\ref{tableBL322} and $j'=1$ in the model in Table~\ref{tableBL321}.
$y^{\Delta}$ Yukawa matrices have entries as 
\begin{align}
y^{\Delta_0}= \left(
\begin{array}{ccc}
 y^{\Delta_0}_{11} & 0 & y^{\Delta_0}_{13}  \\
 0 & 0  & 0  \\
 y^{\Delta_0}_{31} & 0 & y^{\Delta_0}_{33}  \\
\end{array}\right) , 
\qquad
y^{\Delta_1}= \left(
\begin{array}{ccc}
 0 &  y^{\Delta_1}_{12}  & 0  \\
 0 &  y^{\Delta_1}_{22}  & 0  \\
 0 &  y^{\Delta_1}_{32}  & 0  \\
\end{array}\right) ,
\end{align}
 for instance for $X=B_3-L_2$.
We note that we may replace $\Delta_1$ with $\Delta_{-1}$.
Then, the resultant neutrino mass is expressed as
\begin{align}
m_{\nu}= -\frac{(y^{D}_{j{j'}})^2 v^2}{M_{\nu_R^{j'}}}
 + y^{\Delta_0} v_{\Delta}^0+ y^{\Delta_1} v_{\Delta}^1
\end{align}
 where $M_{\nu_R^{j'}} = y^{\nu_R}_{j'} v_2/\sqrt{2}$, and $v_{\Delta}^0 (v_{\Delta}^1)$ are the Majorana mass of $\nu_R^{j'}$, and the VEV $\Delta_0 (\Delta_1) $, respectively.
The first term represents the neutrino mass generated with $j'$th RH neutrino by type-I seesaw mechanism,
 but only one component is generated.
Thus, the mainly triplet Higgs fields have to generate the neutrino mass as above.
The charge of Higgs fields are summarized in Table~\ref{table:Higgs}.
\begin{table}[hbt]
\begin{center}
\begin{tabular}{|c|ccc|c|}
\hline
      &  SU(3)$_c$  & SU(2)$_L$ & U(1)$_Y$ & U(1)$_{B_i-L_j}$  \\ 
\hline
$\Phi$         & {\bf 1 }    &  {\bf 2}       & $ 1/2$    & $0 $   \\  
$\Delta_0$         & {\bf 1 }    &  {\bf 3}       & $ 1$    & $0 $   \\  
$\Delta_1$         & {\bf 1 }    &  {\bf 3}       & $ 1$    & $1 $   \\  
$\phi_2$       & {\bf 1 }    &  {\bf 1}       & $ 0$      & $ + 2 $  \\ 
\hline
\end{tabular}
\end{center}
\caption{The minimal Higgs sector.}
\label{table:Higgs}
\end{table}

In the rest of this paper, we assume that neutrino masses are generated by
 mostly the type-II seesaw mechanism as mentioned above.
Here we comments on another possibility with type-I seesaw mechanism with RH neutrinos, instead of type-II seesaw.
To form Dirac neutrino masses and generate Majorana masses, we need to introduce other Higgs doublets with the charge $({\bf 1 },{\bf 2}, 1/2, \pm x_N )$ and another $U(1)_X$ charged scalar $\phi_{2x_N}$ with the charge $({\bf 1 },{\bf 1}, 0, 2 x_N )$.
In this case, the scalar spectrum includes physical Nambu-Goldstone modes, for example, charged massless scalars, whose existence is excluded by experiments.
Hence, we need further extensions.
As we have seen, unless we introduce additional RH neutrinos with a vanishing $U(1)$ charge,
 we need complicated extensions of Higgs sector to avoid unwanted Nambu-Goldstone modes.

\section{Anomalous magnetic moment of muon}
\label{Sec:g-2}

\subsection{$Z'$ boson contribution to muon $g-2$ }

As in the $L_{\mu}-L_{\tau}$ model~\cite{Baek:2001kca}, 
this $B_3-L_2$ model can also reconcile the muon $g-2$ problem through the $Z'$ boson loop contribution, since the muon is charged under the extra $U(1)$.
By comparing the SM prediction with the latest experimental result~\cite{Muong-2:2021ojo}, the discrepancy is given as 
\begin{equation}
\Delta a_{\mu} = (25.1 \pm 5.9)\times 10^{-10},
\end{equation}
 with $a_{\mu} \equiv (g_{\mu}-2)/2$.
The new contribution from $Z'$ boson loop corrections is estimated as~\cite{Baek:2001kca}
\begin{align}
 a_{\mu}= \frac{g_X^2}{4\pi}\int_0^1dx\frac{2m_{\mu}^2x^2(1-x)}{x^2 m_{\mu}^2+(1-x)m_{Z'}^2} ,
\end{align}
with
\begin{align}
 m_{Z'}^2= \frac{1}{2}g_X^2v_S^2 ,
\end{align}
where we have omitted a negligible kinetic mixing contribution to $m_{Z'}^2$.

\subsection{Constraints from neutrino trident processes}

As in the $L_{\mu}-L_{\tau}$ model, the constraint from neutrino trident processes by the CCFR experiment~\cite{CCFR:1991lpl} is imposed on this model.
The coupling to the muon in our model is same as that in the usual $U(1)_{L_{\mu}-L_{\tau}}$ model,
 we quote the bound from Ref.~\cite{Altmannshofer:2019zhy}.

\section{Dirac right-handed neutrino dark matter}
\label{Sec:DM}

A remarkable result of the anomaly cancellation condition we found for the $U(1)_{B_i-L_j}$ model
 is that non-$j$ generations of RH neutrinos can have nonvanishing opposite charges $x_N$ and $-x_N$. 
From now on, for concreteness, we consider $U(1)_{B_3-L_2}$ models listed in Tables~\ref{tableBL322} and \ref{tableBL321}.
Those two RH neutrinos can form the Dirac spinor
\begin{align}
 \chi = \left(
\begin{array}{c}
\nu_R^3 \\
\nu_R^i{}^* 
\end{array}
\right),
\end{align}
 where $\nu_R^i (i=1, \,\mathrm{or}\,\, 2)$ has the $U(1)_{B_3-L_2}$ charge $-x_N$, and may have the Dirac mass
\begin{align}
\mathcal{L} \supset -m \nu_R^i{}^T \epsilon \nu_R^3  + \mathrm{H.c.} = -\overline{\chi}m \chi .
\end{align}
The Lagrangian of the $\chi$ part is read as
\begin{align}
\mathcal{L} = \overline{\chi}\left[ i\gamma^{\mu}\left( \partial_{\mu}- i x_N g_X \left(\frac{s_{\theta}}{c_{\epsilon}}Z_{\mu} +\frac{c_{\theta}}{c_{\epsilon}}Z'_{\mu}\right) \right)-m \right] \chi .
\end{align}
It is worth noting that $\chi$ has no direct coupling with the SM particles thanks to its $U(1)_X$ charge assignment, and this fact guarantees the stability of $\chi$. The DM candidate $\chi$ does not couple with scalar particle either since its mass does not come from the VEV of a scalar field. 
These properties are in a remarkable contrast with RH neutrino DM 
 in the minimal $U(1)_X$ model with the standard charge assignment,
 where the extra $Z_2$ parity has to be introduced by hand
 to stabilize the DM particle and the scalar exchange processes are important for DM physics~\cite{Okada:2010wd,Okada:2016gsh,Okada:2016tci,Seto:2016pks}.\footnote{For a review on this class of models, see e.g., Ref.~\cite{Okada:2018ktp}.}

\begin{figure}[thb]
\centering
\includegraphics[width=0.7\textwidth]{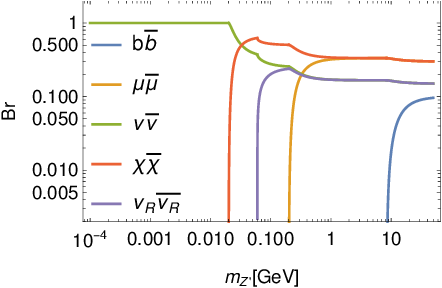}
\caption{The decay branching ratio of the $Z'$ boson for $x_N=1$, $m_{\nu_R} = 0.03$ GeV and $m_{\chi}=10$ MeV.  }
\label{Fig:mXBr}
\end{figure}

For $U(1)_X = U(1)_{B_3-L_2}$ and $x_H=0$, we obtain the decay rate of the $Z'$ gauge boson as
\begin{align}
 \Gamma_{Z'} = & \frac{g_X^2 m_{Z'}}{72 \pi } \left(2 \left( 1-\frac{4 m_t^2}{m_{Z'}^2}\right)^{3/2} +2 \left( 1- \frac{4m_b^2}{m_{Z'}^2}\right)^{3/2} +6 \left(1-\frac{4 m_{\mu}^2}{m_{Z'}^2}\right)^{3/2} +3 \right. \nonumber \\ 
 & \left. +6 x_N^2 \left( 1-\frac{4 m_{\chi}^2}{m_{Z'}^2}\right)^{3/2}+3 \left( 1-\frac{m_{
\nu_R}^2}{m_{Z'}^2}\right) \sqrt{ 1-\frac{4 m_{\nu_R}^2}{m_{Z'}^2}}\right) +\mathcal{O}(\epsilon) ,
\end{align}
 where $m_t, m_b, m_{\mu}$ and $m_{\nu_R}$ are the mass of top, bottom quarks, muon, one RH neutrino with the charge $-1$, and $m_{\chi}$ is the Dirac mass for the other two RH neutrinos. The typical decay branching ratio is shown in Fig.~\ref{Fig:mXBr}.
%

\subsection{Dark matter abundance}

\begin{figure}[thb]
\centering
\includegraphics[width=0.7\textwidth]{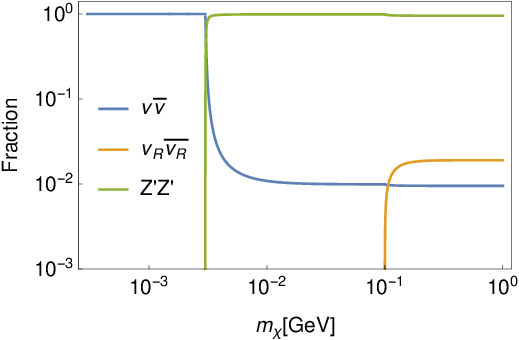}
\caption{An example of the fraction of each annihilation modes for $x_N=5$, $m_{\nu_R} = 0.1$ GeV, and $m_{Z'} = 3\times 10^{-3}$ GeV. }
\label{Fig:mNfr}
\end{figure}
%
\begin{figure}[hbt]
	\begin{center}
		\begin{tabular}{cc}
			\includegraphics[width=0.49\textwidth]{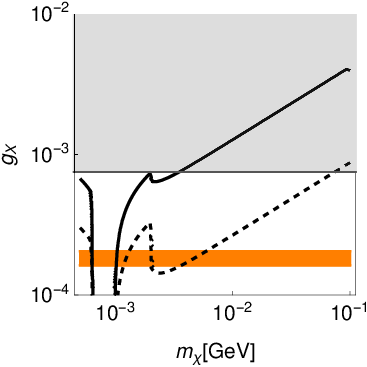}
			 &
			\includegraphics[width=0.48\textwidth]{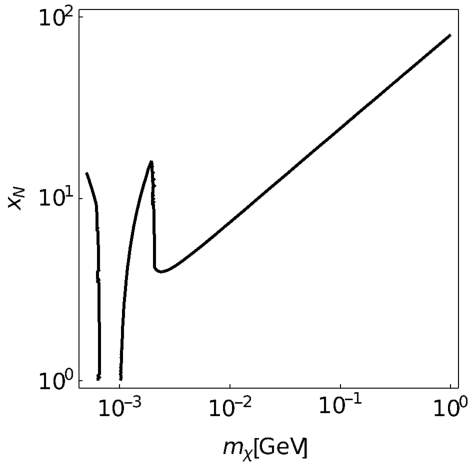}
		\end{tabular}
	\end{center}
\caption{The contour reproducing DM abundance $\Omega_\chi h^2 \simeq 0.1$, for $m_{Z'} = 2$ MeV and $m_{\nu_R} = 0.1$ GeV. 
\textrm{Left}: the case with $x_N=5 (1)$ for the solid (dashed) curve on the $m_{\chi}-g_X$ plane. The gray shaded region is excluded by the neutrino trident constraint~\cite{Altmannshofer:2019zhy}.
\textrm{Right}: the case with $g_X=1.8 \times 10^{-4}$ on the $m_{\chi}-x_N$ plane.   }
\label{Fig:mXgX}
\end{figure}

\begin{figure}[thb]
\centering
\includegraphics[width=0.7\textwidth]{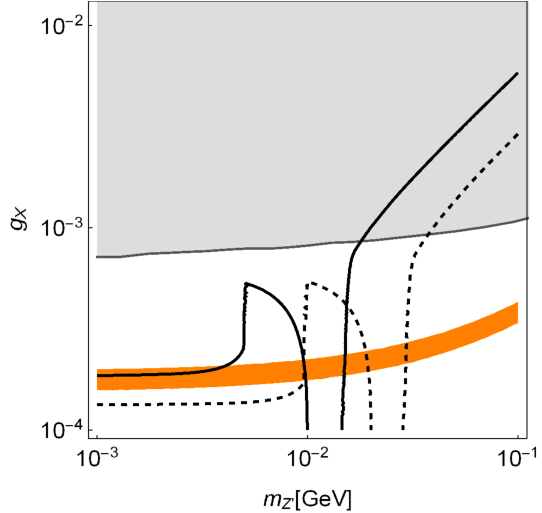}
\caption{The contour reproducing DM abundance $\Omega_{\chi} h^2 \simeq 0.1$ with the muon $g-2$ favored region shaded with orange. The gray shaded region is excluded by the neutrino trident constraint~\cite{Altmannshofer:2019zhy}.
We take $m_{\nu_R} = 0.1$ GeV. The solid (dashed) curve corresponds to $m_{\chi}=5 (10)$ MeV and $x_N=5 (10)$.  }
\label{Fig:mXgXwithg2}
\end{figure}

We estimate the thermal relic abundance of the Dirac DM $\chi$ by solving the Boltzmann equation,
\begin{equation}
 \frac{d n }{dt}+3H n =-\langle\sigma v\rangle ( n^2 - n_{\rm EQ}^2),
\label{eq:boltzman}
\end{equation}
 where $n$ is the number density of $\chi$, $n_{\rm EQ}$ is its number density at thermal equilibrium, $\langle\sigma v\rangle$ is the thermal averaged products of the annihilation cross section and the relative velocity. 
The annihilation channels are $\chi\chi$ to $f\bar{f}$ via $s$-channel $Z'$ exchange and $\chi\chi$ to $Z' Z'$ via $t(u)$-channel $\chi$ exchange.
The later is dominant if the channel is kinematically open as shown in Fig.~\ref{Fig:mNfr}.
The resultant DM relic abundance is given by
\begin{align}
\Omega_{\chi}h^2 = \frac{1.1 \times 10^9 x_d \mathrm{GeV}^{-1}}{\sqrt{ 8\pi g_*}M_P \langle\sigma v\rangle },
\end{align}
where $x_d = m_{\chi}/T_d$ with the decoupling temperature $T_d$~\cite{Kolb:1990vq}.

The contours reproducing the observed DM abundance $\Omega_{\chi}h^2\simeq 0.1$ are shown in Fig.~\ref{Fig:mXgX}.
We note that the constraints that DM dominantly annihilating into muons
 with the mass of the order of GeV is stringently constrained
 from indirect DM searches~\cite{MAGIC:2016xys,Hambye:2019tjt}.
Thus, for $m_{\chi} \gtrsim 100$ MeV, the $Z'$ mass must be smaller than
 the twice of muon mass so that the $Z'$ boson does not decay dominantly into muons. 
The orange strip indicates the parameter region that can solve the muon $g-2$ problem.

In Fig.~\ref{Fig:mXgXwithg2}, we overlay the parameter region to explain the discrepancy
 of the muon $g-2$ on the DM abundance contours. 
This shows that, for example, a set of $x_N=5, m_{\chi}=5$ MeV and $m_{Z'}\lesssim $ several MeV
 or the vicinity of $Z'$ resonance pole ($2 m_{\chi} \sim m_{Z'}$) is able to simultaneously explain the DM abundance and the muon $g-2$ discrepancy.


\section{Other constraints}
\label{Sec:Constraints}

\subsection{Electron neutrino elastic scattering}

The electron neutrino elastic scattering is an effective processes to probe a new interaction~\cite{Harnik:2012ni,Bilmis:2015lja,Lindner:2018kjo,Chakraborty:2021apc,Asai:2023xxl}.
In our model, if the gauge kinetic mixing parameter $\epsilon$ is not vanishing, the electron neutrino scattering is mediated by not only the SM interaction but also new $U(1)_{B_i-L_j}$ gauge interaction. As pointed out in Ref.~\cite{Bilmis:2015lja}, we note the importance of the interference between the SM processes and the $Z'$ boson process.
The relevant part of Lagrangian is
\begin{align}
\mathcal{L} \supset e \hat{A}_{\mu} J_{\hat{A}}^{\mu} + g_2 ( W^+_{\mu} J_{W^+}^{\mu}+ W^-_{\mu} J_{W^-}^{\mu}+ \hat{Z}_{\mu} J_{\hat{Z}}^{\mu}) + g_X \hat{X}_{\mu} J_{\hat{X}}^{\mu} ,
\end{align}
with the currents 
\begin{align}
J_{\hat{A}}^{\mu} =& \sum_k \overline{e^k}\gamma^{\mu} (-1) e^k , \\ 
J_{W^+}^{\mu} =& \sum_k \frac{1}{\sqrt{2}} \overline{\nu^k}\gamma^{\mu} P_L e^k , \\ 
J_{W^-}^{\mu} =& \sum_k \frac{1}{\sqrt{2}} \overline{e^k}\gamma^{\mu} P_L \nu^k  , \\ 
J_{\hat{Z}}^{\mu} = & \sum_k \frac{1}{c_W}\left[ \overline{\nu^k}\gamma^{\mu}\left(\frac{1}{2}\right)P_L\nu^k
  + \overline{e^k}\gamma^{\mu}\left( \left(-\frac{1}{2}+s_W^2\right)P_L + s_W^2 P_R \right) e^k \right], \\
J_{\hat{X}}^{\mu} =& \overline{\nu^j}\gamma^{\mu} (-1)P_L \nu^j+\overline{e^j}\gamma^{\mu} (-1) e^j . 
\end{align}
This can be recast for the mass eigenstates of gauge bosons as
\begin{align}
\mathcal{L}_{\mathrm{int}} = &  
 \frac{g_2}{\sqrt{2}}\sum_k\left( W^+_{\mu}   \overline{\nu^k}\gamma^{\mu} P_L e^k + W^-_{\mu} \overline{e^k}\gamma^{\mu} P_L \nu^k \right) \nonumber \\
& +e (A+c_Ws_{\theta}t_{\epsilon} Z + c_W c_{\theta}t_{\epsilon} Z')\sum_k \overline{e^k}\gamma^{\mu} (-1) e^k \nonumber  \\ 
& +g_2((c_{\theta}-s_{\theta}s_W t_{\epsilon})Z+(-c_{\theta}s_Wt_{\epsilon}-s_{\theta})Z')\sum_k\frac{1}{c_W}\left[  \overline{e^k}\gamma^{\mu}\left( \left(-\frac{1}{2}+s_W^2\right)P_L + s_W^2 P_R \right) e^k \right] \nonumber \\
& + g_X\left(\frac{s_{\theta}}{c_{\epsilon}}Z+\frac{c_{\theta}}{c_{\epsilon}}Z'\right)
  \left[ \overline{e^j}\gamma^{\mu} (-1) e^j + \overline{\nu^j}\gamma^{\mu} (-1)P_L \nu^j \right] \nonumber \\
& +g_2((c_{\theta}-s_{\theta}s_W t_{\epsilon})Z+(-c_{\theta}s_Wt_{\epsilon}-s_{\theta})Z')\sum_k\frac{1}{c_W}\left[ \overline{\nu^k}\gamma^{\mu}\left(\frac{1}{2}\right)P_L\nu^k \right] .
\end{align}
The interaction with $Z'$ boson is expressed as
\begin{align}
\mathcal{L}_{Z'-\mathrm{int}} =& e c_W c_{\theta}t_{\epsilon} Z'_{\mu}\sum_k \overline{e^k}\gamma^{\mu} (-1) e^k \nonumber \\
   &  +\sum_k Z'_{\mu}\overline{\nu^k} \gamma^{\mu} \left[ g_X\frac{c_{\theta}}{c_{\epsilon}}
   (-1) \delta_{kj} +(-c_{\theta}s_Wt_{\epsilon}-s_{\theta})\frac{g_2}{c_W}\left(\frac{1}{2}\right) \right]P_L\nu^k  \nonumber \\
   =& \sum_k g_{e^k Z'} Z'_{\mu} \overline{e^k}\gamma^{\mu} e^k + \sum_k g_{\nu_L^k Z'} Z'_{\mu}\overline{\nu^k} \gamma^{\mu}  P_L\nu^k .
\end{align}
Here, we have defined the effective coupling $g_{e^k Z'}$ and $g_{\nu_L^k Z'}$ as above.
The differential cross section of $k$th flavor (anti)neutrino scattering with electron is given as
\begin{align}
\frac{d \sigma_{\nu_k (\overline{\nu_k})}}{d E_r} = \frac{1}{64\pi E_{\nu}^2 m_e} \overline{|\mathcal{M}|^2} ,
\end{align}
 and the explicit forms are calculated to be
\begin{align}
& \frac{d\sigma(e \nu_e\rightarrow e \nu_e)}{d E_r} \nonumber \\ 
 =& \frac{m_e}{4\pi E_{\nu}^2} \left(2 G_F^2 \left(E_{\nu}^2 \left(2 s_W^2+1\right)^2+4 s_W^4 (E_r-E_{\nu})^2 -2 E_r m_e \left(2 s_W^2+1\right) s_W^2 \right) \right. \nonumber \\ 
 & \left. +\frac{\sqrt{2} G_F g_{\nu_L^k Z'} g_{e^k Z'}  \left(2 E_{\nu}^2+4 s_W^2 \left(2 E_{\nu}^2+E_r^2-E_r (2 E_{\nu}+m_e)\right)-E_r m_e\right)}{\left(2E_r m_e+m_{Z'}^2\right)}  \right. \nonumber \\ 
 & \left.  + \frac{g_{\nu_L^k Z'}^2 g_{e^k Z'}^2\left( 2E_{\nu}^2+E_r^2-E_r(2 E_{\nu}+m_e) \right)}{\left(2E_r m_e+m_{Z'}^2\right)^2} \right) , \\
& \frac{d\sigma(e \nu_{\alpha}\rightarrow e \nu_{\alpha})}{d E_r}  \nonumber \\ 
 =& \frac{m_e}{4\pi E_{\nu}^2} \left(2 G_F^2 \left( E_{\nu}^2 \left(2 s_W^2-1\right)^2 + 4 s_W^4 \left(E_r-E_{\nu}\right)^2  +2 E_r m_e \left(2 s_W^2-1\right) s_W^2 \right) \right. \nonumber \\ 
 & +\frac{\sqrt{2} G_F g_{\nu_L^k Z'} g_{e^k Z'} \left(-2E_{\nu}^2+4s_W^2\left(2E_{\nu}^2+E_r^2-E_r(2 E_{\nu}+m_e)\right)+E_r m_e\right)}{2 E_r m_e+m_{Z'}^2}
  \nonumber \\ & \left.
 +\frac{g_{\nu_L^k Z'}^2 g_{e^k Z'}^2 \left(2E_{\nu}^2+E_r^2-E_r(2E_{\nu}+m_e)\right)}{\left(2E_r m_e+m_{Z'}^2\right)^2}\right), \\
& \frac{d\sigma(e \overline{\nu_{\alpha}}\rightarrow e \overline{\nu_{\alpha}})}{d E_r}  \nonumber \\ 
 =& \frac{m_e}{4\pi E_{\nu}^2} \left(2 G_F^2 \left(4 E_{\nu}^2 s_W^4+\left(1-2 s_W^2\right)^2 (E_r-E_{\nu})^2+2 E_r m_e \left(1-2 s_W^2\right) s_W^2\right)
 \right. \nonumber \\ 
 & +\frac{\sqrt{2} G_F g_{\nu_L^k Z'}g_{e^k Z'}\left(-2(E_{\nu}-E_r)^2+4s_W^2\left(2E_{\nu}^2+E_r^2-E_r(2E_{\nu}+m_e) \right)+E_r m_e \right)}{2 E_r m_e+m_{Z'}^2}
 \nonumber \\ & \left.
 +\frac{g_{\nu_L^k Z'}^2 g_{e^k Z'}^2 \left(2E_{\nu}^2+E_r^2-E_r(2 E_{\nu}+m_e)\right)}{\left(2 E_r m_e+m_{Z'}^2\right)^2}\right),
\end{align}
where $\alpha=\mu$ or $\tau$.

\subsubsection{Borexino constraints}

The differential event rate with respect to the recoil energy $E_r$ at the Borexino detector~\cite{Kumaran:2021lvv} is given by
\begin{align}
\frac{d R}{d E_r} = N_T \int_{E_{\nu}^{\mathrm{min}}}^{\infty} \frac{d\Phi}{dE_{\nu}} \left( P_{ee}\frac{d\sigma_{\nu_e}}{d E_r}+P_{e\mu}\frac{d\sigma_{\nu_{\mu}}}{d E_r}+P_{e\tau}\frac{d\sigma_{\nu_{\tau}}}{d E_r}\right) dE_{\nu} ,
\end{align}
 where $\Phi$ is the flux of solar neutrino, $E_{\nu}$ is the incoming neutrino energy,
\begin{align}
E_{\nu}^{\mathrm{min}} = \frac{1}{2}\left( E_r+\sqrt{E_r^2+2 E_r m_T} \right),
\end{align}
is the minimal neutrino energy to generate the recoil energy $E_r$ by collision with the target with the mass $m_T$, $N_T$ is the number of target particles, and neutrino oscillation effects are taken into account by multiplying the oscillation probability $P_{ij} \equiv P(\nu_i\rightarrow \nu_j)$ for each flavor~\cite{Bahcall:2004mz,Altmannshofer:2019zhy,Nunokawa:2006ms}. The solar neutrino flux are taken from Ref.~\cite{Haxton:2012wfz}.

\begin{figure}[thb]
\centering
\includegraphics[width=0.7\textwidth]{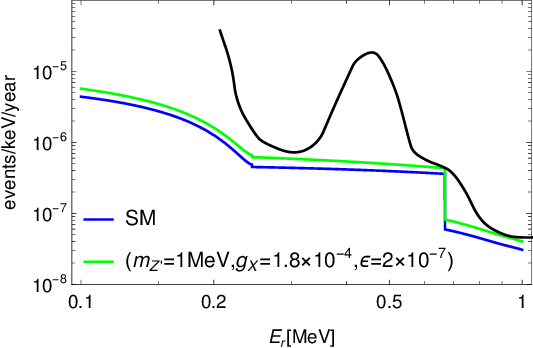}
\caption{The event rate of $e \nu$ scattering as the function of the recoil energy. The black curve is the latest Borexino bound.
The predictions of the SM and of the $U(1)_{B_3-L_2}$ with parameters for the muon $g-2$ and DM explanation are drawn with blue and green curves, respectively. }
\label{Fig:ErrCount}
\end{figure}

We show the theoretical prediction of models along with the Borexino results~\cite{Borexino:2017rsf} (black curve) in Fig.~\ref{Fig:ErrCount}.
For the benchmark point $(g_X=1.8\times 10^{-4}, m_{Z'}=1~\mathrm{MeV})$, which can explain the muon $g-2$ anomaly as well as the thermal DM, we calculate the event rate of $e\nu$ scattering by varying the gauge kinetic mixing parameter $\epsilon$.
As shown by the green line in Fig.~\ref{Fig:ErrCount}, we find the Borexino bound on the gauge mixing parameter to be $|\epsilon| \lesssim \mathcal{O}(10^{-7})$.
For the comparison, the SM prediction is drawn with the blue curve.

\subsubsection{CHARM-II constraint}

\begin{figure}[hbt]
	\begin{center}
		\begin{tabular}{cc}
			\includegraphics[width=0.49\textwidth]{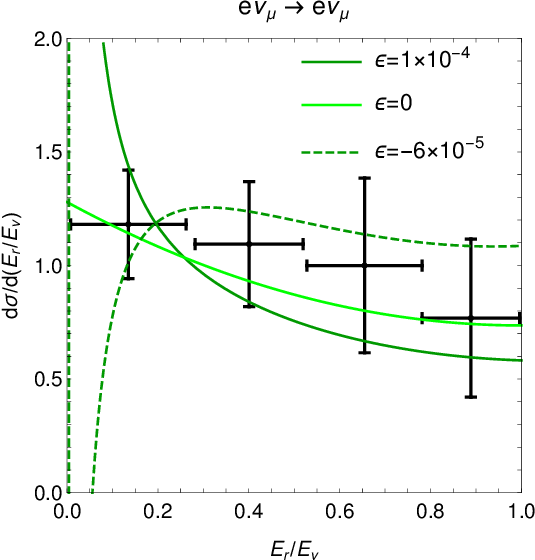}
			 &
			\includegraphics[width=0.49\textwidth]{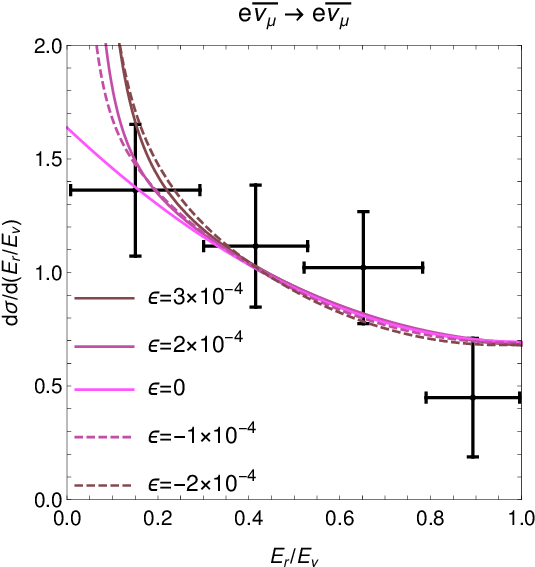}
		\end{tabular}
	\end{center}
\caption{The CHARM-II measurement of the differential cross section as the function of $E_r/E_{\nu}$ and the prediction for the $U(1)_{B_3-L_2}$ model with the muon $g-2$ and DM favored points for various $\epsilon$. 
Left:  $e^- \nu_{\mu} $ scattering.   
Right: $e^- \overline{\nu_{\mu}} $ scattering.}
\label{Fig:charm2}
\end{figure}

Since the extra gauge boson of the $U(1)_{B_3-L_2}$ gauge theory couples with muon neutrinos
 than other flavor of neutrinos, experiments on $e\nu_{\mu}$ or $e\overline{\nu_{\mu}}$ scattering such as CHARM-II~\cite{CHARM-II:1993phx,CHARM-II:1994dzw} would also provide a constraint on our model.
In Fig.~\ref{Fig:charm2}, we show the $U(1)_{B_3-L_2}$ model prediction of the differential cross section for various $\epsilon$, which is compared with the CHARM-II results~\cite{CHARM-II:1993phx}. 
We find, for the benchmark point $(g_X=1.8\times 10^{-4}, m_{Z'}=1 \mathrm{MeV})$, the CHARM-II bound on the gauge kinetic mixing as $-6\times 10^{-5} \lesssim \epsilon \lesssim 2\times 10^{-4}$
 which is less stringent than the Borexino bound shown in the previous subsection.

\section{Summary}
\label{Sec:Summary} 

We have proposed a variation of flavor dependent gauged $U(1)$ extension of the SM.
Motivated by the fact that the gauged $U(1)_{B-L}$ symmetry is anomaly free for each generation,
 an $i$th generation of quarks and $j$th generation of leptons are charged in the $U(1)_{B_i-L_j}$ model.
One generation of RH neutrinos must be charged under the $U(1)_{B_i-L_j}$ for the theory to be the anomaly free. 
There is another nontrivial aspect of the model that the other RH neutrinos may also be charged under the symmetry
 with the nonvanishing opposite charge $x_N$, and hence they form a Dirac neutrino $\chi$. 
This Dirac fermion $\chi$ is stable due to the $U(1)$ gauge invariance and a natural candidate for DM.

Among various possibilities of charge assignments, the $U(1)_{B_3-L_2}$ model is attractive,
 as it may explain the discrepancy of the muon $g-2$ between the experimental results and the SM prediction, and, in addition, the LHC constraints are relatively weak because the $Z'$ interacts with only third generation of quarks. 
We have shown that, in a certain parameter region, the muon $g-2$ anomaly and the thermal DM abundance are simultaneously explained
 without contradicting other experimental bounds if the gauge kinetic mixing is small enough.


\section*{Acknowledgments}
This work is supported in part by the U.S. DOE Grant No.~DE-SC0012447 (N.O.), 
 the Japan Society for the Promotion of Science (JSPS) KAKENHI Grants
 No.~19K03860, No.~19K03865, and No.~23K03402 (O.S.).




\begin{thebibliography}{99}

\bibitem{Pati:1973uk}
  J.~C.~Pati and A.~Salam,
  Phys. Rev. D \textbf{8}, 1240-1251 (1973).
\bibitem{Davidson:1978pm}
  A.~Davidson,
  Phys. Rev. D \textbf{20}, 776 (1979).
\bibitem{Mohapatra:1980qe}
  R.~N.~Mohapatra and R.~E.~Marshak,
  Phys.\ Rev.\ Lett.\  {\bf 44}, 1316 (1980)
  [Erratum-ibid.\  {\bf 44}, 1644 (1980)].
\bibitem{Mohapatra:1980}
  R.~E.~Marshak and R.~N.~Mohapatra,
  Phys.\ Lett.\  B {\bf 91}, 222 (1980).

\bibitem{Babu:2017olk} 
  K.~S.~Babu, A.~Friedland, P.~A.~N.~Machado and I.~Mocioiu,
  JHEP {\bf 1712}, 096 (2017).
\bibitem{Alonso:2017uky} 
  R.~Alonso, P.~Cox, C.~Han and T.~T.~Yanagida,
  Phys.\ Lett.\ B {\bf 774}, 643 (2017).
\bibitem{Bian:2017rpg} 
  L.~Bian, S.~M.~Choi, Y.~J.~Kang and H.~M.~Lee,
  Phys.\ Rev.\ D {\bf 96}, no. 7, 075038 (2017).
\bibitem{Cox:2017rgn} 
  P.~Cox, C.~Han and T.~T.~Yanagida,
  JCAP {\bf 1801}, no. 01, 029 (2018).
\bibitem{Elahi:2019drj}
F.~Elahi and A.~Martin,
Phys. Rev. D \textbf{100}, no.3, 035016 (2019).
\bibitem{Okada:2019sbb}
N.~Okada and O.~Seto,
Phys. Rev. D \textbf{101}, no.2, 023522 (2020).


\bibitem{He:1990pn}
X.~G.~He, G.~C.~Joshi, H.~Lew and R.~R.~Volkas,
Phys. Rev. D \textbf{43}, R22-24 (1991).
\bibitem{Foot:1990mn}
R.~Foot,
Mod. Phys. Lett. A \textbf{6}, 527-530 (1991).
\bibitem{Baek:2001kca}
S.~Baek, N.~G.~Deshpande, X.~G.~He and P.~Ko,
Phys. Rev. D \textbf{64}, 055006 (2001).
\bibitem{Ma:2001md}
E.~Ma, D.~P.~Roy and S.~Roy,
Phys. Lett. B \textbf{525}, 101-106 (2002).

\bibitem{Chun:2018ibr}
E.~J.~Chun, A.~Das, J.~Kim and J.~Kim,
JHEP \textbf{02}, 093 (2019)
[erratum: JHEP \textbf{07}, 024 (2019)].
\bibitem{Greljo:2021xmg}
A.~Greljo, P.~Stangl and A.~E.~Thomsen,
Phys. Lett. B \textbf{820}, 136554 (2021).
\bibitem{Barman:2019aku}
B.~Barman, D.~Borah, P.~Ghosh and A.~K.~Saha,
JHEP \textbf{10}, 275 (2019).
\bibitem{Wang:2019byi}
W.~Wang and Z.~L.~Han,
Phys. Rev. D \textbf{101}, no.11, 115040 (2020).
\bibitem{Barman:2021yaz}
B.~Barman, P.~Ghosh, A.~Ghoshal and L.~Mukherjee,
JCAP \textbf{08}, no.08, 049 (2022).
\bibitem{Crivellin:2015lwa}
A.~Crivellin, G.~D'Ambrosio and J.~Heeck,
Phys. Rev. D \textbf{91}, no.7, 075006 (2015).
\bibitem{Bonilla:2017lsq}
C.~Bonilla, T.~Modak, R.~Srivastava and J.~W.~F.~Valle,
Phys. Rev. D \textbf{98}, no.9, 095002 (2018).
\bibitem{Ko:2017quv}
P.~Ko, T.~Nomura and H.~Okada,
Phys. Lett. B \textbf{772}, 547-552 (2017).
\bibitem{Allanach:2020kss}
B.~C.~Allanach,
Eur. Phys. J. C \textbf{81}, no.1, 56 (2021)
[erratum: Eur. Phys. J. C \textbf{81}, no.4, 321 (2021)].
\bibitem{Allanach:2022blr}
B.~Allanach and E.~Loisa,
JHEP \textbf{03}, 253 (2023).
\bibitem{Ban:2021tos}
K.~Ban, Y.~Jho, Y.~Kwon, S.~C.~Park, S.~Park and P.~Y.~Tseng,
PTEP \textbf{2023}, no.1, 013B01 (2023).

\bibitem{LHCb:2022qnv}
 [LHCb],
[arXiv:2212.09152 [hep-ex]].   

\bibitem{Aoyama:2020ynm}
T.~Aoyama, N.~Asmussen, M.~Benayoun, J.~Bijnens, T.~Blum, M.~Bruno, I.~Caprini, C.~M.~Carloni Calame, M.~C\`e and G.~Colangelo, \textit{et al.}
Phys. Rept. \textbf{887}, 1-166 (2020).


\bibitem{Appelquist:2002mw}
  T.~Appelquist, B.~A.~Dobrescu and A.~R.~Hopper,
  Phys.\ Rev.\ D {\bf 68} 035012 (2003).
\bibitem{Oda:2015gna}
S.~Oda, N.~Okada and D.~s.~Takahashi,
Phys. Rev. D \textbf{92}, no.1, 015026 (2015).
\bibitem{Das:2016zue}
A.~Das, S.~Oda, N.~Okada and D.~s.~Takahashi,
Phys. Rev. D \textbf{93}, no.11, 115038 (2016).

\bibitem{Cho:2020mnc}
W.~Cho, K.~Y.~Choi and S.~M.~Yoo,
Phys. Rev. D \textbf{102}, no.9, 095010 (2020).

\bibitem{Minkowski:1977sc}
P.~Minkowski,
Phys. Lett. B \textbf{67}, 421-428 (1977).
\bibitem{Yanagida:1979as}
T.~Yanagida,
Conf. Proc. C \textbf{7902131}, 95-99 (1979).
\bibitem{GellMann:1980vs}
M.~Gell-Mann, P.~Ramond and R.~Slansky,
Conf. Proc. C \textbf{790927}, 315-321 (1979).
\bibitem{Mohapatra:1979ia}
R.~N.~Mohapatra and G.~Senjanovic,
Phys. Rev. Lett. \textbf{44}, 912 (1980).

\bibitem{Schechter:1980gr}
J.~Schechter and J.~W.~F.~Valle,
Phys. Rev. D \textbf{22}, 2227 (1980).
\bibitem{Magg:1980ut}
M.~Magg and C.~Wetterich,
Phys. Lett. B \textbf{94}, 61-64 (1980).
\bibitem{Cheng:1980qt}
T.~P.~Cheng and L.~F.~Li,
Phys. Rev. D \textbf{22}, 2860 (1980).

\bibitem{Muong-2:2021ojo}
B.~Abi \textit{et al.} [Muon g-2],
Phys. Rev. Lett. \textbf{126}, no.14, 141801 (2021).

\bibitem{CCFR:1991lpl}
S.~R.~Mishra, S.~A.~Rabinowitz, C.~Arroyo, K.~T.~Bachmann, R.~E.~Blair, C.~Foudas \textit{et al.} [CCFR],
Phys. Rev. Lett. \textbf{66}, 3117-3120 (1991).
\bibitem{Altmannshofer:2019zhy}
W.~Altmannshofer, S.~Gori, J.~Mart\'\i{}n-Albo, A.~Sousa and M.~Wallbank,
Phys. Rev. D \textbf{100}, no.11, 115029 (2019).


\bibitem{Okada:2010wd}
N.~Okada and O.~Seto,
Phys. Rev. D \textbf{82}, 023507 (2010).
\bibitem{Okada:2016gsh}
N.~Okada and S.~Okada,
Phys. Rev. D \textbf{93}, no.7, 075003 (2016).
\bibitem{Okada:2016tci}
N.~Okada and S.~Okada,
Phys. Rev. D \textbf{95}, no.3, 035025 (2017).
\bibitem{Seto:2016pks}
O.~Seto and T.~Shimomura,
Phys. Rev. D \textbf{95}, no.9, 095032 (2017).
\bibitem{Okada:2018ktp}
S.~Okada,
Adv. High Energy Phys. \textbf{2018}, 5340935 (2018).

\bibitem{Kolb:1990vq}
E.~W.~Kolb and M.~S.~Turner, {\it The Early Universe}, Addison-Wesley (1990).

\bibitem{MAGIC:2016xys}
M.~L.~Ahnen \textit{et al.} [MAGIC and Fermi-LAT],
JCAP \textbf{02}, 039 (2016).
\bibitem{Hambye:2019tjt}
T.~Hambye and L.~Vanderheyden,
JCAP \textbf{05}, 001 (2020).

\bibitem{Harnik:2012ni}
R.~Harnik, J.~Kopp and P.~A.~N.~Machado,
JCAP \textbf{07}, 026 (2012).
\bibitem{Bilmis:2015lja}
S.~Bilmis, I.~Turan, T.~M.~Aliev, M.~Deniz, L.~Singh and H.~T.~Wong,
Phys. Rev. D \textbf{92}, no.3, 033009 (2015).
\bibitem{Lindner:2018kjo}
M.~Lindner, F.~S.~Queiroz, W.~Rodejohann and X.~J.~Xu,
JHEP \textbf{05}, 098 (2018).
\bibitem{Chakraborty:2021apc}
K.~Chakraborty, A.~Das, S.~Goswami and S.~Roy,
JHEP \textbf{04}, 008 (2022).
\bibitem{Asai:2023xxl}
K.~Asai, A.~Das, J.~Li, T.~Nomura and O.~Seto,
[arXiv:2307.09737 [hep-ph]].


\bibitem{Kumaran:2021lvv}
S.~Kumaran, L.~Ludhova, \"O.~Penek and G.~Settanta,
Universe \textbf{7}, no.7, 231 (2021).

\bibitem{Bahcall:2004mz}
J.~N.~Bahcall and C.~Pena-Garay,
New J. Phys. \textbf{6}, 63 (2004).

\bibitem{Nunokawa:2006ms}
H.~Nunokawa, S.~J.~Parke and R.~Zukanovich Funchal,
Phys. Rev. D \textbf{74}, 013006 (2006).

\bibitem{Haxton:2012wfz}
W.~C.~Haxton, R.~G.~Hamish Robertson and A.~M.~Serenelli,
Ann. Rev. Astron. Astrophys. \textbf{51}, 21-61 (2013).

\bibitem{Borexino:2017rsf}
M.~Agostini \textit{et al.} [Borexino],
Phys. Rev. D \textbf{100}, no.8, 082004 (2019).

\bibitem{CHARM-II:1993phx}
P.~Vilain \textit{et al.} [CHARM-II],
Phys. Lett. B \textbf{302}, 351-355 (1993).
\bibitem{CHARM-II:1994dzw}
P.~Vilain \textit{et al.} [CHARM-II],
Phys. Lett. B \textbf{335}, 246-252 (1994).


\end{thebibliography}
\end{document}